\documentclass[twocolumns, 11pt]{IEEEtran}

\usepackage{floatflt}

\usepackage{color}
\usepackage{cite}

\usepackage{graphicx}
\usepackage{amsmath}
\usepackage{amsthm}
\usepackage{amssymb}

\usepackage{verbatim}

\usepackage{psfrag,array}

\usepackage{subfigure}

\usepackage{stfloats}

\usepackage{todonotes}

\usepackage{amsmath}
\usepackage{epstopdf}
\usepackage{algorithmic}

\newcommand{\p}[1]{\mathop{\mbox{\it p} } }

\renewcommand{\vec}[1]{\ensuremath{\boldsymbol{#1}}}
\newcommand{\be}{\begin{equation}}
\newcommand{\ee}{\end{equation}}
\newcommand{\ba}{\begin{array}}
\newcommand{\ea}{\end{array}}
\newcommand{\bea}{\begin{eqnarray}}
\newcommand{\eea}{\end{eqnarray}}
\newcommand{\bean}{\begin{eqnarray*}}
\newcommand{\eean}{\end{eqnarray*}}
\newcommand{\argmax}{\mathop{\arg\max}}
\newcommand{\argmin}{\mathop{\arg\min}}

\newcommand{\rmh}{^{\rm \dag}}

\definecolor{white}{rgb}{1,1,1}

\begin{document}

\title{Deep-Neural-Network based Fall-back Mechanism in Interference-Aware Receiver Design}

\author
{
\begin{tabular}{c}
 Sha Hu, Dzevdan Kapetanovic, Neng Wang, and Wenquan Hu
 \thanks{The authors are with Huawei Research Center in Lund, Sweden (\{firstname.lastname\}@huawei.com).}
\end{tabular}
\vspace*{-4mm}
}

\maketitle

\begin{abstract}
In this letter we consider designing a fall-back mechanism in an interference-aware receiver. Typically, there are two different manners of dealing with interference, known as enhanced interference-rejection-combining (eIRC) and symbol-level interference-cancellation (SLIC). Although SLIC performs better than eIRC, it has higher complexity and requires the knowledge of modulation-format (MF) of interference. Due to potential errors in MF detection, SLIC can run with a wrong MF and render limited gains. Therefore, designing a fall-back mechanism is of interest that only activates SLIC when the detected MF is reliable. Otherwise, a fall-back happens and the receiver turns to eIRC. Finding a closed-form expression of an optimal fall-back mechanism seems difficult, and we utilize deep-neural-network (DNN) to design it which is shown to be effective and performs better than a traditional Bayes-risk based design in terms of reducing error-rate and saving computational-cost.
\end{abstract}


\section{Introduction}
In modern communication systems, multiple users and spatial data-streams are multiplexed on the same time-frequency resources to increase spectral-efficiency. This brings challenge when recovering useful data from superimposed received signals. In design of interference-aware receiver, two approaches are widely applied, namely, enhanced interference-rejection-combining (eIRC) and symbol-level interference-cancellation (SLIC)~\cite{3gpp}. SLIC detects and reconstructs interference, and then remove it from the received signal before detecting useful signal, while eIRC suppresses interference instead of canceling it~\cite{GD15}. SLIC is more advanced, but its computational cost and processing latency are also higher. 

Typically, eIRC receivers only require the knowledge of interference channel, or even simpler, the covariance matrix of interference. Such information can be obtained based on estimating the channel with pilots or correlating the received samples. For SLIC receivers, except for channel information, modulation format (MF) is also needed in order to decode and reconstruct the interference. MF classification is a classical problem and a rich literature can be found in e.g.,~\cite{DK12, BC16, XG17}.

One issue that has not yet been widely considered in literature, to our best knowledge, is adaptively switching between SLIC and eIRC, which we refer to as a ''fall-back mechanism'' in an interference-aware receiver design. The necessity of fall-back lies in the fact that it is more beneficial to run eIRC instead of SLIC when MF detection is unreliable. Since SLIC yields a high computational cost and the gains is insignificant in these cases compared to MF. Although performance-wise eIRC receiver is suboptimal, it is however, much simpler and robust against MF mismatch.

In this letter, we consider two different designs of fall-back mechanism: a traditional Bayes-risk minimization based~\cite{K98}, and a deep-neural-network (DNN) based. As finding an optimal fall-back strategy seems out of reach, DNN is used to find a suitable mechanism through training. As we show in this letter, the DNN based design have advantages compared to the Bayes-risk approach in reducing error-rate and saving computational-cost, which makes it a promising feature in solving similar problems in communication systems.

\section{Interference-Aware Receiver Design}
The multiple-input multiple-output (MIMO) model with interference is modeled as
\bea \label{sm1}   \vec{y} =\vec{H}\vec{s}+\vec{G}\vec{x}+\vec{n}, \eea
where $\vec{H}$ and $\vec{G}$ are effective MIMO channels (including possible power-offsets and precoding matrices \cite{3gpp})  of desired signal and interference, respectively, which are known at the receiver. The transmitted symbol vector $\vec{s}$ is yet to be detected and interfered by $\vec{x}$. Further, the noise $\vec{n}$ comprises i.i.d. complex-valued AWGN variables with zero-mean and a variance $N_0$.

Note that the signal model (\ref{sm1}) can also represent a multi-user MIMO (MU-MIMO) transmission written as
\bea \label{sm2}   \vec{y} =\left[ {\begin{array}{cc} \!\vec{H}  & \vec{G}  \!\end{array} } \right]  \left[ {\begin{array}{c} \!\vec{s}  \\ \vec{x} \! \end{array} } \right]+\vec{n}. \eea

\subsection{eIRC Receiver}

An eIRC receiver estimates $\vec{s}$ via LMMSE filtering as
\bea \label{eIRC}   \tilde{\vec{s}} =\vec{H}\rmh\left(\vec{H}\vec{H}\rmh+\vec{G}\vec{G}\rmh+N_0\vec{I}\right)^{-1}\vec{y}, \eea
where $\vec{H}\rmh$ is the Hermitian of $\vec{H}$. In a simpler form (i.e., IRC receiver), it uses
\bea \label{IRC}     \tilde{\vec{s}}=\vec{H}\rmh\left(\vec{H}\vec{H}\rmh+\vec{R}\right)^{-1}\vec{y}, \eea
where $\vec{R}$ is the estimated covariance matrix of interference plus noise.

\subsection{SLIC Receiver}
With SLIC, the interference and its covariance matrix are estimated (e.g. LMMSE) as $\hat{\vec{x}}$ and $\vec{C}_{\vec{x}}$. Then, $\hat{\vec{x}}$ is removed from $\vec{y}$, yielding
\bea \label{sm3}   \hat{\vec{y}} =\vec{H}\vec{s}+\vec{G}\big(\vec{x}-\hat{\vec{x}}\big)+\vec{n}. \eea
The signal $\vec{s}$ is estimated with purified samples as
\bea \label{SLIC}    \tilde{\vec{s}} =\vec{H}\rmh\left(\vec{H}\vec{H}\rmh+\vec{G}\vec{C}_{\vec{x}}\vec{G}\rmh+N_0\vec{I}\right)^{-1}\hat{\vec{y}} .\eea
In order to apply SLIC, the MF of interference $\vec{x}$ is needed (while the MF of $\vec{s}$ is known). With an advanced MIMO detector, instead of using LMMSE (\ref{SLIC}), a joint detection of $(\vec{s},\vec{x})$ can also be carried out on (\ref{sm2}) via sphere-decoding (SD) or its variants. 

\subsection{MF Detection with ML }
Optimal MF detection of $\vec{x}$ is maximum-likelihood (ML) \cite{DK12}, which can be formulated as
\bea  \label{optML}   \mathcal{X}_2^{\text{ML}} &\!\!\!\!\!\!\!\!\!\!=\!\!\!\!\!\!\!\!\!\!&\argmin\limits_{\mathcal{X}_2}\bigg(\sum_{k=0}^{K-1}\!\sum_{\substack{\vec{s}_k\in \mathcal{X}_1^{K_1},\\ \vec{x}_k\in \mathcal{X}_2^{K_2}}}\!\!\!\| \vec{y}_k \!-\!\vec{H}_k\vec{s}_k\!-\!\vec{G}_k\vec{x}_k\|^2 \notag \\ &&\!+\!KN_0\big(K_1\ln|\mathcal{X}_1|+K_2\ln|\mathcal{X}_2|\big)\bigg),\eea
where $\mathcal{X}_1$ and $\mathcal{X}_2$ denote MFs used for $\vec{s}_k$ and $\vec{x}_k$, respectively, and without loss of generality, we let $K_1$ layers in $\vec{s}_k$ and $K_2$ layers in $\vec{x}_k$ use identical MFs. The parameter $K$ is the number of samples used for detection.
 
Computing (\ref{optML}) directly renders a prohibitive complexity since the size of summation is equal to $|\mathcal{X}_1|^{K_1}|\mathcal{X}_2|^{K_2}$. A simplified detection can be
\bea  \label{subML}   \tilde{\mathcal{X}}_2 =\argmax\limits_{\mathcal{X}_2}\mu(\mathcal{X}_2),\eea
where the probability-metric of each MF is computed with $\tilde{K}\!=\!K_2K$ interference estimates as
\bea \label{prob} \mu(\mathcal{X}_2)\!=\!-\sum_{k=0}^{\tilde{K}-1}\!\sum_{x_{k}\in \mathcal{X}_2}\!\!\big(\| \hat{x}_k -x_k\|^2 +\tilde{N}_{k}\ln|\mathcal{X}_2|\big).\eea
The estimate $\hat{x}_k$ on the $k$th layer of $\vec{x}$ is obtained from
\bea   \label{hatx1} \hat{\vec{x}} =\vec{G}\rmh\left(\vec{H}\vec{H}\rmh+\vec{G}\vec{G}\rmh+N_0\vec{I}\right)^{-1}\vec{y}, \eea
with $\tilde{N}_{k}$ being the effective noise associated with it. To improve the accuracy of $\hat{\vec{x}}$, the signal $\vec{s}$ can be estimated as $\hat{\vec{s}}$ first, and removed from $\vec{y}$ before estimating $\vec{x}$, which yields
\bea   \label{hatx2} \hat{\vec{x}} =\vec{G}\rmh\!\left(\vec{H}\vec{C}_{\vec{s}}\vec{H}\rmh+\vec{G}\vec{G}\rmh+N_0\vec{I}\right)^{-1}\!\big(\vec{y}-\vec{H}\hat{\vec{s}}\big). \eea

MF detection with (\ref{subML}) has much lower computational-cost compared to the original ML in (\ref{optML}). A further simplification to (\ref{prob}) is proposed in \cite{DK12} as
\be  \mu(\mathcal{X}_2)\!\approx\!-\sum_{k=0}^{K_2-1}\!\!\big(\min_{x_k\in \mathcal{X}_2}\big(\| \hat{x}_k -x_k\|^2\! +\!f(\tilde{N}_{k}, \mathcal{X}_2)\big),\ee
where $f(\tilde{N}_{k}, \mathcal{X}_2)$ denotes a look-up-table (LUT) operation that depends on $\tilde{N}_{k}$ and $\mathcal{X}_2$.

\begin{figure}[t]
\begin{center}
\vspace{-1mm}
\hspace*{2mm}
\scalebox{.315}{\includegraphics{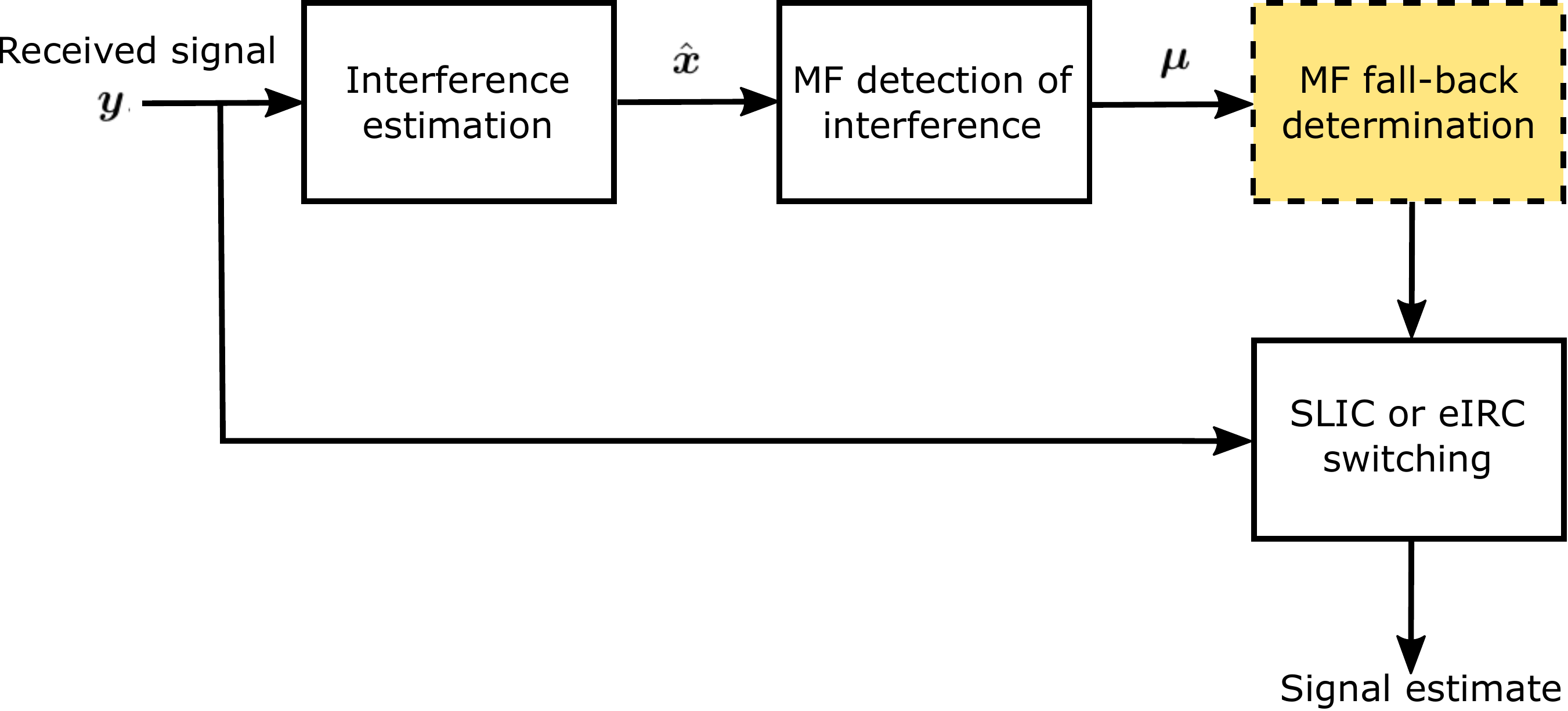}}
\vspace{-6mm}
\caption{\label{fig1}Interference-aware receiver design with MF fall-back.}
\vspace{-7mm}
\end{center}
\end{figure}

\section{Fall-back Mechanism Designs}
One potential issue with interference-aware receiver design is that ML detection of interference MF has a certain error-probability. Due to a high complexity of SLIC, it is beneficial to run eIRC when the gain of SLIC is limited, especially under cases that MF decisions are unreliable. Therefore, a fall-back mechanism is useful to adaptively switch the receiver between SLIC and eIRC.

Assuming ML detection of interference MF in (\ref{subML}) is applied and we obtain a probability vector from (\ref{prob}) as 
\be \vec{\mu}\!=\!\frac{1}{\sum\limits_{m=0}^{M-1}\exp(\mu_m)}\big(\exp(\mu_0), \exp(\mu_1), \dots, \exp(\mu_{M-1})\big),  \ee
corresponding to each of the $M$ MFs. Our objective of designing a fall-back mechanism is to determine when the ML decision of interference MF is reliable so that SLIC (\ref{SLIC}) is activated, otherwise a fall-back occurs and eIRC (\ref{eIRC}) is used to save complexity. This is equivalent to specify a decision-boundary on a hyperplane of the probability distribution function (pdf) $p(\vec{\mu})$ depicted in~Fig.~\ref{fig2}.

\subsection{Problem Formulation}
\begin{figure}[t]
\begin{center}
\vspace{-2mm}
\hspace*{2mm}
\scalebox{.42}{\includegraphics{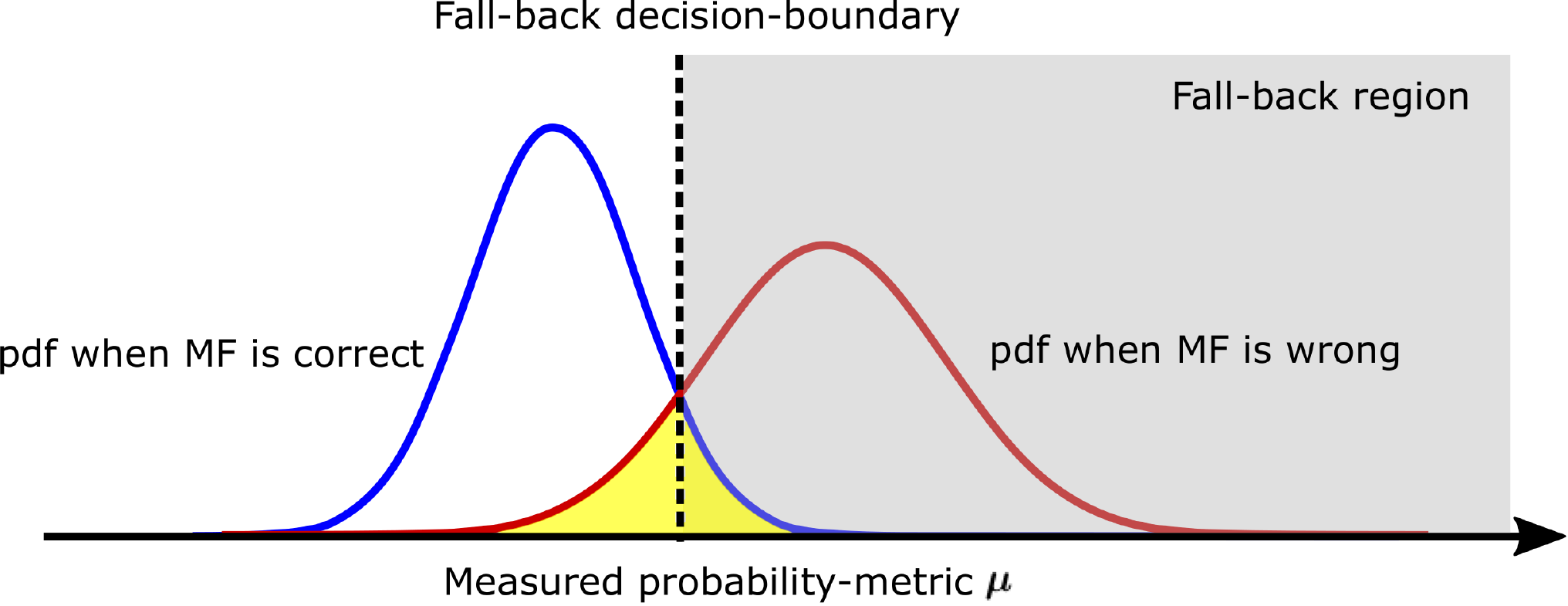}}
\vspace*{-8mm}
\caption{\label{fig2}Optimal fall-back boundary that minimizes $R_e$ in (\ref{err-rate}).}
\vspace{-7mm}
\end{center}
\end{figure}

Denoting correct and error probabilities of interference MF detection as $P_d$ and $P_e$, respectively, to formulate the design problem we define an error-rate $R_e$ as
\bea \label{err-rate} R_e&\!\!\!\!=\!\!\!\!&P_d\gamma_{e,1}+P_e \gamma_{e,2}.  \eea
where
\bea \gamma_{e,1}&\!\!\!\!=\!\!\!\!&P\big(\text{Fall-back\big|Interference MF is correct}\big), \notag \\
 \gamma_{e,2}&\!\!\!\!=\!\!\!\!&P\big(\text{Not fall-back\big|Interference MF is wrong}\big)\! .  \qquad \eea
With ideal fall-back (unreachable since the pdf curves of $\vec{p(\vec{\mu})}$ when MF decision is correct and wrong are typically overlapped as shown in Fig. \ref{fig2}), it holds that $R_e\!=\! \gamma_{e,1}\!=\! \gamma_{e,2}\!=\!0$, whereas without fall-back it holds that $R_e\!=\!P_e$. The error probabilities $ \gamma_{e,1}$ and $ \gamma_{e,2}$ are determined by the fall-back mechanism applied, and shall be minimized to reduce $R_e$ .

\subsection{Bayes-Risk based Fall-back Design}

Denoting the hypothesis $\mathcal{H}_m$ as transmitting $\vec{x}$ with the $m$th MF, one approach to design fall-back is based on minimizing a general Bayes-risk defined as
\bea \label{bayes} B_e=\sum_{n=0}^{M}\sum_{m=0}^{M-1}c_{m,n}P(\mathcal{H}_n|\mathcal{H}_m)P(\mathcal{H}_m),\eea
where $P(\mathcal{H}_m)$ is the probability of hypothesis $\mathcal{H}_m$, and $P(\mathcal{H}_n|\mathcal{H}_m)$ is the conditional probability of deciding hypothesis $\mathcal{H}_n$ under true hypothesis $\mathcal{H}_m$, associated with a cost-weights $c_{m,n}$. It holds that $c_{m,n}\!\geq\!0$ and $c_{m,m}\!=\!0$ for any $m$ and $n$. Note that in (\ref{bayes}) we have added an extra hypothesis $\mathcal{H}_M$ for denoting a fall-back decision, although it does not exist in a true hypothesis. 

Denoting decision region $\mathcal{R}_n\!=\!\{\hat{\vec{x}};\; \text{deciding as}\; \mathcal{H}_n\}$, we can rewrite (\ref{bayes}) as
\bea \label{bayes1} B_e&\!\!\!\!=\!\!\!\!&\sum_{n=0}^{M}\sum_{m=0}^{M-1}\int_{\mathcal{R}_n}c_{m,n}P(\hat{\vec{x}} |\mathcal{H}_m)P(\mathcal{H}_m)\mathrm{d}\hat{\vec{x}} \notag \\
&\!\!\!\!=\!\!\!\!&\sum_{n=0}^{M}\sum_{m=0}^{M-1}\int_{\mathcal{R}_n}c_{m,n}P(\mathcal{H}_m|\hat{\vec{x}} )P(\hat{\vec{x}})\mathrm{d}\hat{\vec{x}}
\notag \\
&\!\!\!\!=\!\!\!\!&\sum_{n=0}^{M}\int_{\mathcal{R}_n}\epsilon_n(\hat{\vec{x}} ) P(\hat{\vec{x}} )\mathrm{d}\hat{\vec{x}}
\eea
where 
\bea \label{eps} \epsilon_n(\hat{\vec{x}} )=\sum_{m=0}^{M-1}c_{m,n}P(\mathcal{H}_m|\hat{\vec{x}})\propto\sum_{m=0}^{M-1}c_{m,n}\exp(\mu_m).\eea
Since each observation $\hat{\vec{x}}$ can be assigned to one decision region, it has to be the one that has the smallest $\epsilon_n(\hat{\vec{x}} )$. This leads to a following decision rule:
\bea \label{bayes2} \hat{n}=\argmin\limits_{0\leq n \leq M} \epsilon_n(\hat{\vec{x}} ).\eea
As seen from (\ref{eps}), the cost-weights $c_{m,n}$ determines the decision regions, which is yet to be optimized.

\vspace*{-1mm}
\subsection{DNN based Fall-back Design}

\begin{figure}[t]
\begin{center}
\vspace{-0mm}
\hspace{-2mm}
\scalebox{.33}{\includegraphics{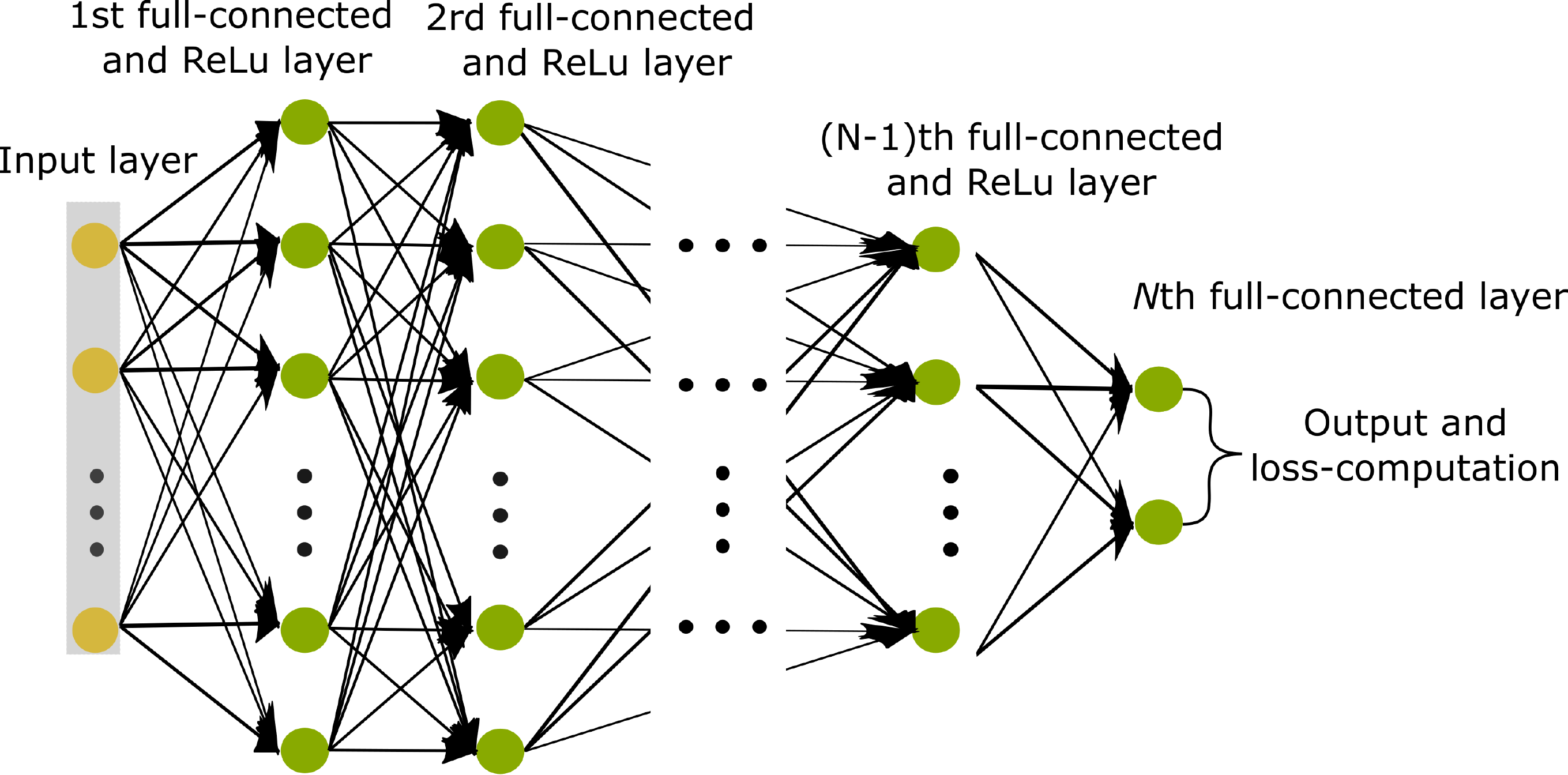}}
\vspace*{-6mm}
\caption{\label{fig3} DNN structure used for receiver fall-back.}
\vspace{-7mm}
\end{center}
\end{figure}

To derive an optimal fall-back mechanism that minimizes $\gamma_{e,1}$ and $ \gamma_{e,2}$, one needs to know the pdf of $\vec{\mu}$ which is $M$-dimensional and difficult. As well known, DNN can be effective in solving problems whose relationships between input and output variables are not explicitly known. We also use DNN to design the fall-back. The input to DNN is $\vec{\mu}$, an $M\!\times\!1$ real-valued vector (typically $M\!=\!4$ comprising QPSK, 16QAM, 64QAM, and 256QAM) with non-negative elements (whose sum is 1). We use $N$ full-connected layers with each associated with a rectified-linear-unit (ReLU) layer (except the last one), which is depicted in Fig. \ref{fig3}.

Recall that in full-connected layer the input-output pair ($\vec{z}_n, \vec{z}_{n+1}$) is computed as
\bea \vec{z}_{n+1}=\vec{A}_n\vec{z}_n+\vec{b}_n. \eea
Followed by ReLu layer, we update $\vec{z}_{n+1}$ as
\bea \label{layer} \hat{\vec{z}}_{n+1}=\max(\vec{z}_{n+1},\vec{0})=\max(\vec{A}_n\vec{z}_n+\vec{b}_n,\vec{0}).\eea
With such a DNN-structure, we can write the operations in a compact form as
\be \label{dnn} \vec{\eta}=\vec{A}_N \big(g_{N-1}\circ\cdots\circ g_2\circ g_{1}(\vec{\mu})\big)+\vec{b}_N, \ee
where $g_{n+1}$ represents the function specified in (\ref{layer}). The final output $\vec{\eta}$ is of size $2\times 1$, whose first element $\eta_0$ represents the probability of not falling-back and the other $\eta_1$ represents the probability of falling-back. 

To train parameters $\vec{w}_i=(\vec{A}_i, \vec{b}_i)$ ($1\leq i\leq N$), a soft-max loss function is used,
\bea \label{loss}  L(\vec{\eta}; \vec{\mu}, \vec{\ell}, \vec{w} )=-\eta_\ell+\ln\big(\exp(\eta_0)+\exp(\eta_1)\big), \eea
where $\ell=0, 1$ are labels indicating MF decision to be correct or wrong (matched with definitions of $\eta_0$ and $\eta_1$). The parameters are optimized using gradient-approach, which is computed recursively with chain-rule as
\bea \label{dev1} \frac{\partial L}{\partial \vec{w}_i}= \frac{\partial L}{\partial \vec{\eta}}\cdot \frac{\partial \vec{\eta}}{\partial \vec{y}_{n-1}} \cdot \frac{\partial \vec{y}_{n-1}}{\partial \vec{y}_{n-2}}\cdots \cdot\frac{\partial \vec{y}_i}{\partial \vec{w}_i}.  \eea
Over training, $\vec{w}_i$ are updated with a learning-rate $\alpha$ as
\bea \vec{w}_i^{n+1}=\vec{w}_i^{n}-\alpha\frac{\partial L}{\partial \vec{w}_i}.  \eea

\section{Simulation Results and Discussions}

\subsection{DNN Training Accuracy}
Theoretically, one needs to minimize the averaged loss
\bea \bar{L}(\vec{\eta};\vec{\mu},\ell_n,\vec{w})=\frac{1}{S}\sum_{n=0}^{S-1} L(\vec{\eta}_n;\vec{\mu}_n,\ell_n,\vec{w}),\eea
for all input $S$ training vectors, where $\vec{w}$ comprises all parameters of the DNN. The input probability-metric vector $\vec{\mu}_n$ and its label $\ell_n$ are obtained from ML detection of interference MF. However, such an optimization is too large and slow when training size $S$ is large, and calculating gradients over the entire sum is also difficult. Hence, we use $16$ training vectors as a batch in one iteration to update $\vec{w}$, and in-total we run $S/16$ iterations. Throughout tests, the learning-rate $\alpha$ is set to 0.01. 

We set $N\!=\!4$ in DNN (4 full-connected layers of sizes (8, 8, 4, 2) and 3 ReLu layers in Fig. \ref{fig3}), and the training vectors $(\vec{\mu}_n, \ell_n)$ are generated under AWGN channel\footnote{In training phase, we model the estimate of interference $\hat{x}$ in (\ref{hatx2}) as true symbols plus AWGN.}.  We use (\ref{hatx2}) to estimate interference with $\tilde{K}\!=\!24$ samples, and then use (\ref{subML}) to estimate MF and the probability-metrics, which are later sent to DNN for determining fall-back. 

\begin{figure}[t]
\begin{center}
\vspace{-1mm}
\hspace{2mm}
\scalebox{.31}{\includegraphics{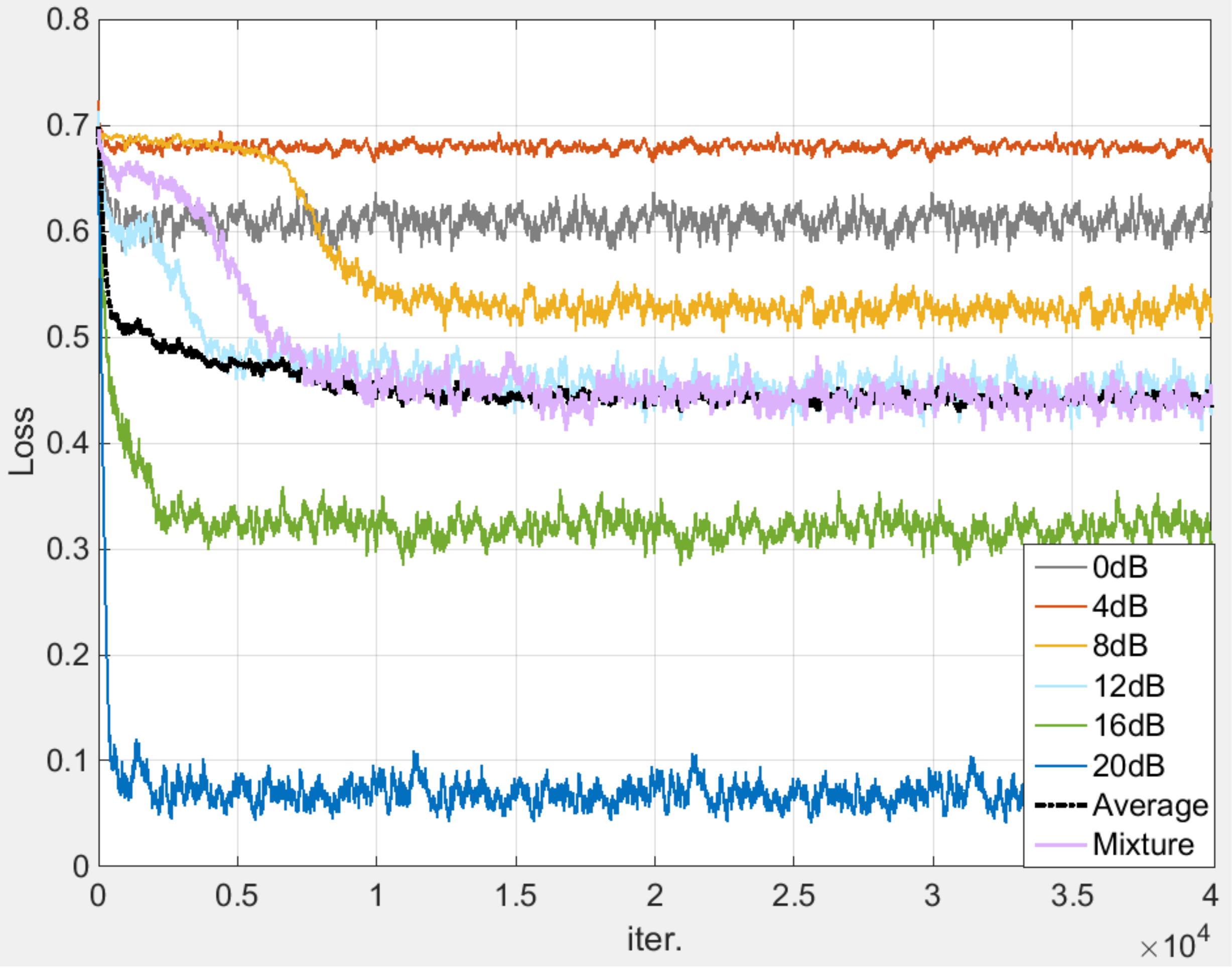}}
\vspace{-2mm}
\caption{\label{fig4} Losses in DNN training.}
\vspace{-8mm}
\end{center}
\end{figure}

The soft-max losses is depicted in Fig. \ref{fig4}, where we showed the losses both for training at a specific SNR and a mixture of different SNRs.  As can be seen, we set $S\!=\!640,000$ for each training, and iteration number is equal to 40,000. When SNR increases from 0 to 20dB, the losses decrease. What is interesting is that the DNN is robust against signal-to-noise ratio (SNR) changes. This can be seen from that training with mixed SNRs (samples from 0, 4, up to 20dB are mixed with a equal probability for training) yields similar losses to the curve with averaged losses over different SNRs. The later one can be seen as an upper-bound for a mixture-training, since the losses are optimized for each SNR point individually before averaging. Moreover, as we have adopted MMSE pre-processing in (\ref{hatx2}), the channel has been equalized which means that the DNN is also robust against channel selectivities. 

\vspace{-3mm}
\subsection{Error-Rate Comparisons}

\begin{figure}[t]
\begin{center}
\vspace{-2mm}
\hspace{-3mm}
\scalebox{.283}{\includegraphics{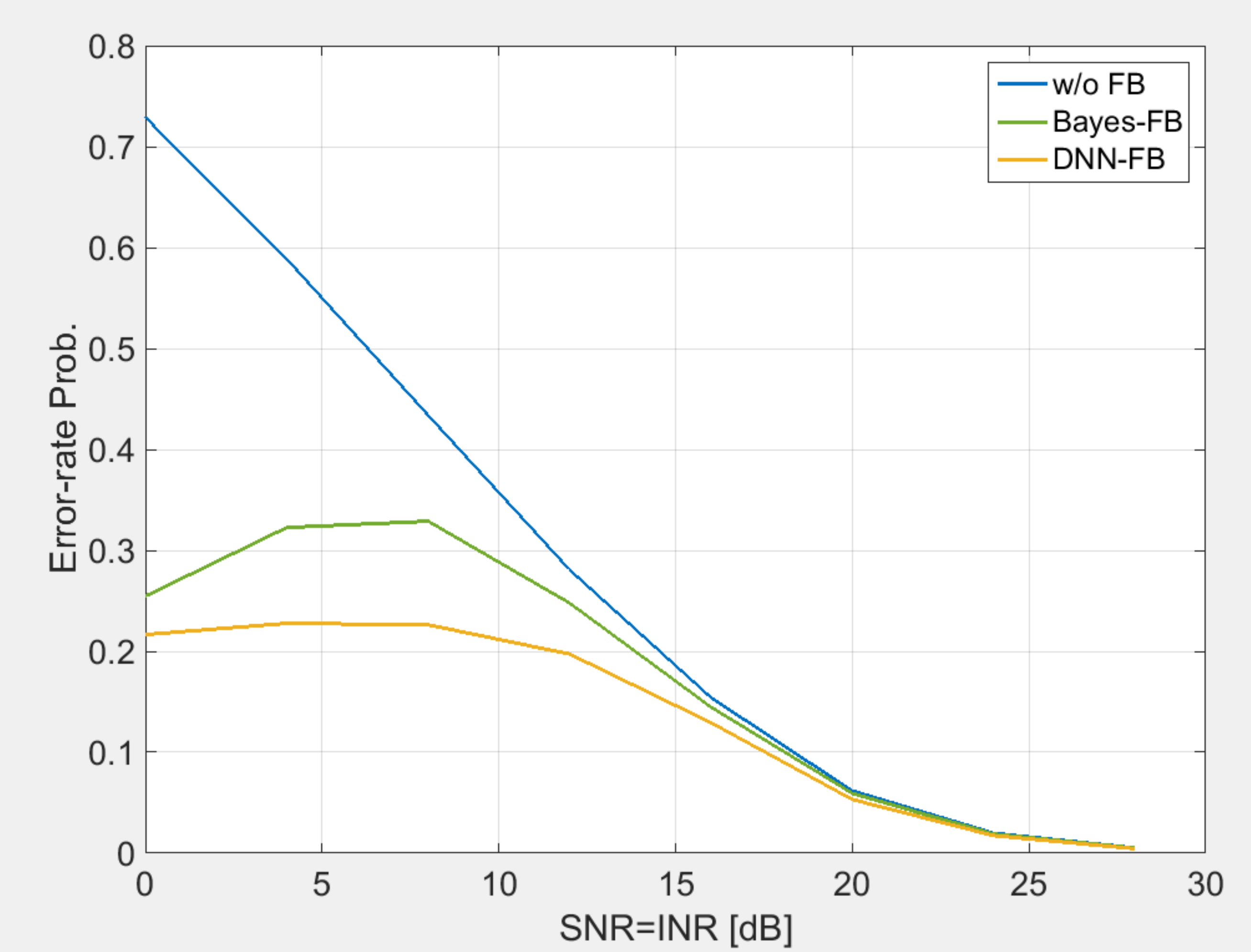}}
\vspace{-3mm}
\caption{\label{fig5} Error-rates evaluations.}
\vspace{-4mm}
\end{center}
\end{figure}

The error-rate probability defined in (\ref{err-rate}) is shown in Fig. \ref{fig5}, where we compare error-rates for cases with no fall-back, Bayes-risk based fall-back, and DNN based fall-back. The coefficients $c_{m,n}$ of Bayes-risk are optimized for minimizing averaged error-rate over different MFs and fixed for all test cases. As can be seen, the DNN based fall-back yields a minimal error-rate, which is averaged when both signal $\vec{s}$ and interference $\vec{x}$ are modulated with QPSK, 16QAM and 64QAM, respectively. The channels $\vec{H}$ and $\vec{G}$ are modeled as Extended-Pedestrian-A (EPA)\cite{3gpp} with 5Hz Doppler shift.

In Fig. \ref{fig6}, we further show the probabilities of correct MF detection and fall-back. Due to fall-back, the correct detection probabilities of Bayes-risk and DNN based fall-backs are degraded compared to the case without fall-back. But the DNN based fall-back is more aggressive in determining a fall-back than the Bayes based approach. This means that on averaged the DNN based fall-back will activate SLIC receiver much less than the other two cases, and yields a higher complexity-saving by running eIRC on more PRBs.

\vspace{-3mm}
\subsection{Complexity Savings}
\begin{figure}
\begin{center}
\vspace{-0mm}
\hspace{-3mm}
\scalebox{.283}{\includegraphics{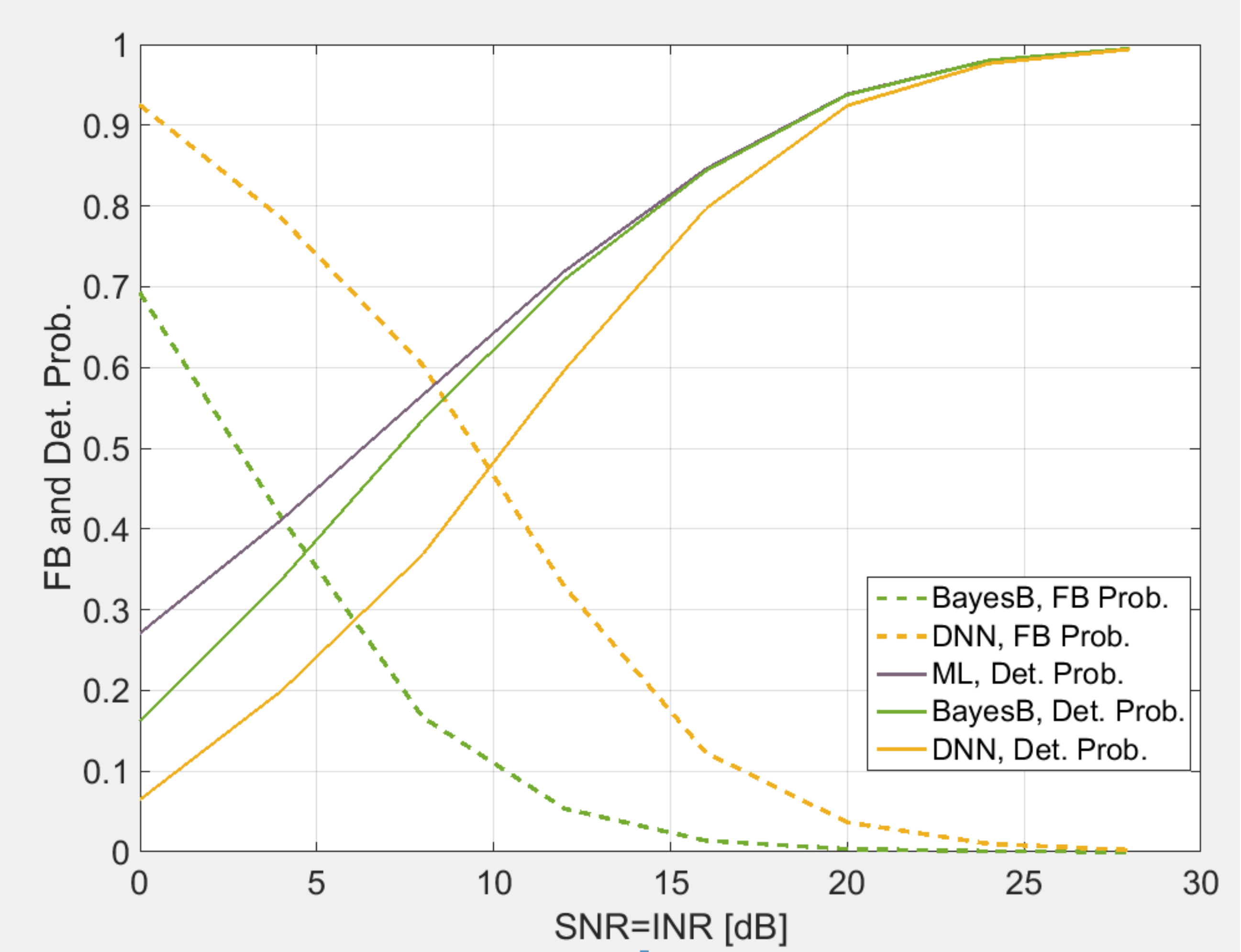}}
\vspace{-2mm}
\caption{\label{fig6} Probabilities of detection and fall-back where SNR and interference-to-noise ratio (INR) are equal.}
\vspace{-8mm}
\end{center}
\end{figure}

The detailed complexity savings by applying the fall-back mechanism depends on the types of SLIC and eIRC receivers. But we can still measure the relative complexity saving by the ratio of activating SLIC and eIRC receivers. With no fall-back, the SLIC is always running which yields the highest complexity. With Bayes-risk based fall-back, seen from Fig. \ref{fig6}, the fall-back probability is 10\% at 10dB SNR, which only saves 10\% of SLIC operations that are replaced them by eIRC. On the other hand, the fall-back probability of DNN based approach is around 50\%, which replaces half SLIC operations by eIRC, and complexity savings are more significant.

Since DNN based fall-back mechanism yields a higher fall-back probability, i.e., less activation of the advanced SLIC receiver, the impact on throughput is of interest.

\vspace{-3mm}
\subsection{Throughput Comparisons}
\begin{figure}[t]
\begin{center}
\vspace{-2mm}
\hspace{2mm}
\scalebox{.255}{\includegraphics{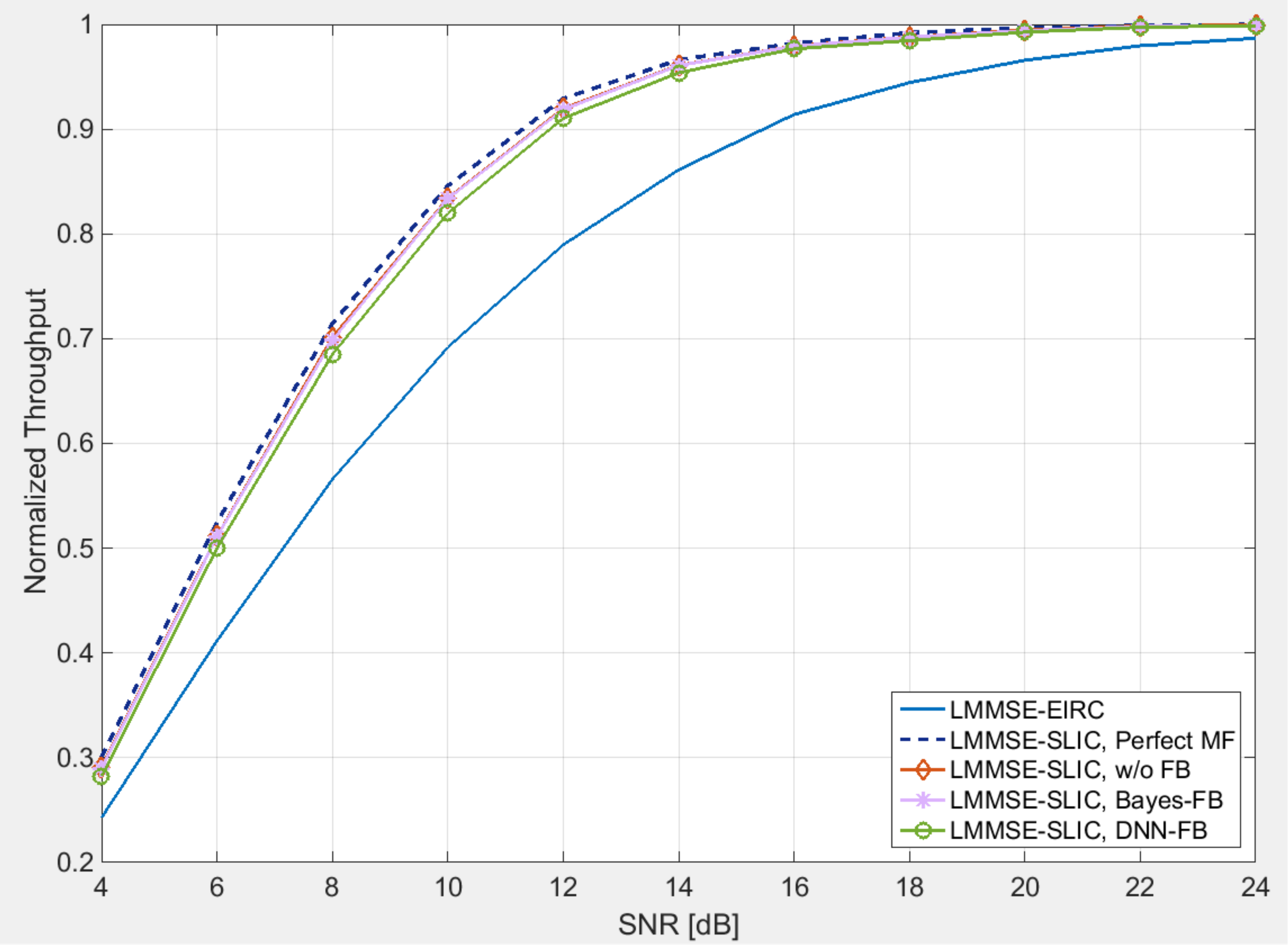}}
\vspace{-2mm}
\caption{\label{fig7} Normalized throughputs.}
\vspace{-6mm}
\end{center}
\end{figure}

In Fig. \ref{fig7} we test the normalized throughput on a 1.4MHz bandwidth in long-term-evolution (LTE) system~\cite{3gpp} under the conditions that $\vec{s}$ is QPSK modulated while $\vec{x}$ uses 16QAM, and with a code-rate 0.5. The MF detection on each physical resource block (PRB) is independent and based on the first 24 samples. 

As can be seen that the gains between eIRC and SLIC with genie MF is close to 4dB at 90\% throughput. Surprisingly, the case without fall-back performs quite close to the ideal throughput, and so are the cases with Bayes and DNN based fall-back mechanisms. However, as the later two approaches activating less SLIC compared to the case without fall-back, the savings in complexity are still siginificant. Especially with the DNN based fall-back, the computational-cost saving is much higher than Bayes based fall-back, as seen from Fig. \ref{fig6}, but throughput losses compared to others is marginal.

\section{Summary}
In this letter we have considered the design of a fall-back mechanism in an interference-aware receiver. Based on MMSE estimate of interference, the MF of it can be detected using ML criterion where the probability-metric for each possible MF has been computed. With these metrics, we have derived a Bayes based fall-back mechanism whose cost-weights have no closed-form expressions. We have then constructed a DNN based fall-back mechanism comprising of full-connected and ReLu layers. As verified by simulations, the DNN is robust against SNR and channel variations which are typical obstacles encountered when applying DNN in baseband processing. Moreover, the DNN based fall-back yields a smaller error-rate than Bayes based fall-back and no fall-back, and a higher fall-back probability which saves more computational-cost than other methods by less activation of SLIC. Furthermore, throughput simulations have showed that the DNN only has marginal throughput loss compared to the case with no fall-back.

\bibliographystyle{IEEEtran}

\end{document}